\documentclass[aps,prb,twocolumn,showpacs,superscriptaddress]{revtex4-1}
\usepackage{graphicx}
\usepackage{dcolumn}
\usepackage{amsmath}
\usepackage{amssymb}
\usepackage{subfigure,amsmath,verbatim,moreverb,bm}
\def\be{\begin{equation}}
\def\ee{\end{equation}}
\def\ber{\begin{eqnarray}}
\def\eer{\end{eqnarray}}

\def\xv{{\bf x}}
\def\rv{{\bf r}}

\def\jv{{\bf j}}

\def\Im{{\rm Im}}
\def\Re{{\rm Re}}
\begin{document}
\title{Time-dependent density functional theory on a lattice}
\author{M. Farzanehpour}
\email{m.farzanehpour@gmail.com}
\affiliation{Nano-Bio Spectroscopy group and ETSF Scientific Development Centre, 
  Departamento de F\'isica de Materiales, Universidad del Pa\'is Vasco UPV/EHU, E-20018 San Sebasti\'an, Spain}

\author{ I. V. Tokatly}
\email{ilya\_tokatly@ehu.es}
\affiliation{Nano-Bio Spectroscopy group and ETSF Scientific Development Centre, 
  Departamento de F\'isica de Materiales, Universidad del Pa\'is Vasco UPV/EHU, E-20018 San Sebasti\'an, Spain}
\affiliation{IKERBASQUE, Basque Foundation for Science, E-48011 Bilbao, Spain}

\date{\today}
\begin{abstract}
A time-dependent density functional theory (TDDFT) for a quantum many-body system on a lattice is formulated rigorously. We prove the uniqueness of the density-to-potential mapping and demonstrate that a given density is $v$-representable if the initial many-body state and the density satisfy certain well defined conditions. In particular, we show that for a system evolving from its ground state any density with a continuous second time derivative is $v$-representable and therefore the lattice TDDFT is guaranteed to exist. The TDDFT existence and uniqueness theorem is valid for any connected lattice, independently of its size, geometry and/or spatial dimensionality. The general statements of the existence theorem are illustrated on a pedagogical exactly solvable example which displays all details and subtleties of the proof in a transparent form. In conclusion we briefly discuss remaining open problems and directions for a future research.
\end{abstract}
\pacs{31.15.ee, 71.10.Fd} 
\maketitle

\section{Introduction}

Time-dependent density functional theory (TDDFT) is rapidly becoming one of the most popular methods for modeling dynamics of realistic quantum many-particle systems. Nowadays TDDFT is routinely applied to dynamical problems in condensed matter, atomic, and nuclear physics, in quantum chemistry, and in nanoscience. Only within last few years several journals published special issues on TDDFT \cite{TDDFT-PCCP,TDDFT-TEOCHEM,Helbigetal2011} [see also a recent review, Ref.~\onlinecite{CasHui2012}] and two TDDFT textbooks came out earlier this year \cite{TDDFT-2012,Ullrich-book}. 

The reasons for the popularity of TDDFT are obviously the same as those for the ground state DFT. It reduces calculations of the density in a complicated interacting many-body system to solving a set of Hartree-type equations for a reference system of noninteracting Kohn-Sham (KS) particles, which simplifies the problem dramatically. A possibility of such a reduction rests on two fundamental mathematical statements: (i) a one-to-one mapping between the density and the external potential, meaning that two different (modulo a constant) potentials cannot produce the same density; and (ii) a $v$-representability of the density, i.~e. the existence of a potential that generates a given density. The first statement, known as the mapping theorem, guarantees that a many-body wave function and thus any observable are unique functionals of the density. The $v$-representability is required to justify the KS construction in (TD)DFT. 

In the static DFT a proof of the mapping theorem was presented in a seminal paper by Hohenberg and Kohn \cite{hohenbergkohn}, while a general solution of a more tricky $v$-representability problem appeared much later and only for lattice systems \cite{Kohn1983,Chayes1985} which is probably sufficient in most practical cases. Proving corresponding theorems for TDDFT turned out to be much more difficult, mostly because of the absence of a minimum principle in dynamics. Only in 1985 Runge and Gross (RG) succeeded to find a sufficiently general proof of the TDDFT mapping theorem for a class of analytic in time ($t$-analytic) potentials \cite{RunGro1984}. An attempt to attack the time-dependent $v$-representability problem has been performed in Ref.~\onlinecite{vanLeeuwen1999} by assuming $t$-analyticity both for potentials and for allowed densities. Under this restriction a formal power series for the potential can be uniquely reconstructed from a given Taylor expansion of the density. Unfortunately the convergence of that series is unproved up to now, and thus a complete solution of the $v$-representability problem within the series expansion technique is still lacking. We note that the issues of $t$-analyticity and a uniform convergence of power series in quantum dynamics are not as exotic as it may appear on the first sight \cite{Holstein1972,Maitra2010,ZenMaiBur2012}. Despite a number of indications\cite{vanLeeuwen2001,RugPenBau2010} that $t$-analyticity was not a fundamental limitation of the theory, a question of a more general and clean justification of TDDFT remained open for many years.

Very recently it has been recognized \cite{Tokatly2007,TokatlyPCCP2009,Maitra2010,Tokatly2011,TokatlyTDCDFT2011,RugLee2011,Ruggenthaleretal2012} that the existence of all TDDFT-type theories is equivalent to the solvability of a certain universal nonlinear many-body problem which determines the potential and the many-body wave function in terms of a given basic observable. Mathematically this universal problem can be posed in two different forms. The first possibility is to view it as a Cauchy problem for a special nonlinear Schr\"odinger equation (NLSE) \cite{Tokatly2007,Maitra2010,TokatlyPCCP2009,Tokatly2011,TokatlyTDCDFT2011}. The uniqueness and the existence of solutions to this NLSE are equivalent, respectively, to the mapping and the $v$-representability theorems in TDDFT. Alternatively it can be formulated as a fixed point problem for a certain nonlinear map in the space of potentials \cite{RugLee2011}. First applications of the above two formulations appeared almost simultaneously in the last year \cite{TokatlyTDCDFT2011,RugLee2011}.

The fixed point approach has been used in Ref.~\onlinecite{RugLee2011} to prove the existence of the ``classical`` TDDFT in its original RG form. In this work the $t$-analyticity requirement was completely relaxed and effectively replaced by a more physical and plausible assumption -- a boundness of a certain generalized response function related to a stress-density correlator. 

A rigorous formulation of a time-dependent current density functional theory (TDCDFT) on a lattice was presented in Ref.~\onlinecite{TokatlyTDCDFT2011} within NLSE formulation of the problem. The lattice TDCDFT turned out to the first and, in fact, the simplest example of a TDDFT-type theory for which both the mapping and the $v$-representability theorems have been proved without any unjustified assumption. 

In the present paper we further develop and extend the lattice NLSE technique of Ref.~\onlinecite{TokatlyTDCDFT2011} to address a long standing problem of TDDFT for lattice many-body systems. It may appear surprising, but even the RG mapping theorem for $t$-analytic potentials is currently absent for the lattice TDDFT. This should be contrasted to the lattice TDCDFT where the standard power series argumentation can be easily adapted \cite{StePerCin2010}. Many discussions of mathematical and conceptual difficulties of the lattice TDDFT can be found in the literature \cite{Verdozzi2008,Baer2008,Kurth2011,vettkarkarver,Verdozzietal2011}. In this work we show how to overcome these problems using NLSE formalism. We prove the uniqueness and existence theorems for the lattice TDDFT and analyze the conditions one has to impose on the initial state and the density to guarantee $v$-representability. In particular we demonstrate that practically any properly normalized density is locally $v$-representable provided the dynamics starts from the ground state. These results put applications of TDDFT to various lattice models \cite{Verdozzi2008,Baer2008,Kurth2011,vettkarkarver,Verdozzietal2011,Kurthetal2010,KarPriVer2011,SteKur2011,BerLiuBur2012,Verdozzietal2011} on a firm ground, and shed new light on the general mathematical structure of TDDFT.

The structure of the paper is the following. In Sec.~II we present a general formulation of the lattice many-body theory and derive a lattice analog of the force balance equation that plays a key role in the NLSE formalism for TDDFT. Section~II is the central part of the present paper. Section~IIIA starts with a brief review of our approach to TDDFT in a more familiar continuum case. Then we derive the corresponding NLSE for a lattice theory, and, finally, formulate and prove the general existence a uniqueness theorem on the lattice TDDFT. Several important aspects of the basic theorem for a generic initial state are discussed in Sec.~IIIB. Section~IV presents an explicit illustration of the general existence theorem for a simple exactly solvable model -- one particle on a two-site lattice. In Sec.~V we consider a practically important case of a system evolving from its ground state. The main outcome of this section is a $v$-representability theorem for the initial ground state. In Conclusion we review our results and discuss open questions and directions for further research. In Appendix we derive an explicit NLSE for another illustrative model describing dynamics of two interacting particles on two sites (a Hubbard dimer).

\section{Preliminaries: Many-body problem on a lattice}

In this work we are considering quantum dynamics of $N$ interacting particles on a lattice which consists of a finite, but arbitrary large number $M$ of sites. The state of the system at time $t$ is characterized by a many-body wave function $\psi(\rv,\rv_2,...,\rv_N;t)$, where coordinates $\rv_i$ of particles $(i=1,2,..,N)$ take values on the lattice sites. The dynamics driven by an external scalar potential $v(\rv;t)$ is described by the following discrete Schr\"odinger equation
\begin{eqnarray} \label{SE1}
i\partial_t \psi(\rv_1,...,\rv_N;t)=&-&\sum_{i=1}^N\sum_{\xv_i} T_{\rv_i ,\xv_i} \psi(...,\xv_i,...;t)\nonumber\\
&+& \sum_{j=1}^N v(\rv_j;t)  \psi(\rv_1,...,\rv_N;t)\nonumber\\
&+& \sum_{j > i} w_{\rv_i , \rv_j}  \psi(\rv_1,...,\rv_N;t),
\end{eqnarray} 
where real coefficients $T_{\rv,\rv'}=T_{\rv',\rv}$ correspond to the rate of hopping from site $\rv$ to the site $\rv'$ (we assume that $T_{\rv,\rv}=0$), and $w_{\rv,\rv'}$ is a potential of a pairwise particle-particle interaction. To cover various possible physical applications we do not specify the geometry of the lattice and do not assume as usual that the interaction depends only on the distance between particles. For example, the latter is important in a typical transport setup with noninteracting or weakly interacting leads connected to a strongly interacting central region \cite{Kurth2005,Kurthetal2010,SteKur2011,BerLiuBur2012,Trosteretal2012}.

Equation \eqref{SE1} (formally it corresponds to a system of $M^N$ linear ordinary differential equations) determines the wave function as a unique functional of the external potential and a given initial state,
\begin{equation}\label{IW}
\psi(\rv_1,\rv_2,...,\rv_N;t_0)=\psi_0(\rv_1,\rv_2,...,\rv_N).
\end{equation} 

The key object of DFT is the density of particles $n(\rv;t)$, which in the present context means the number of particles on a given site
\begin{equation} \label{D1}
 n(\rv;t)= N \sum_{\rv_2,...,\rv_N} \vert \psi(\rv,\rv_2,...,\rv_N;t) \vert ^2,
\end{equation} 
where we assumed that the particles are identical. By taking the time derivative of the definition (\ref{D1}) and using the Schr\"odinger equation \eqref{SE1}, we find the following equation of motion for the density
\begin{equation} \label{CE1}
\dot n(\rv;t)= i \sum_{\rv'}[ T_{\rv,\rv'}\rho(\rv,\rv';t)-T_{\rv',\rv}\rho(\rv',\rv;t)],
\end{equation}
where $\dot n = \partial_t n$, and $\rho(\rv,\rv';t)$ is a density matrix (or a ``link density``) on the $[\rv,\rv']$-link,
\begin{equation} \label{rho1}
\rho(\rv,\rv';t)= N \sum_{\rv_2,...,\rv_N} \psi^*(\rv,\rv_2,...,\rv_N;t)\psi(\rv',\rv_2,...,\rv_N;t).
\end{equation}

Obviously Eq.~(\ref{CE1}) is a lattice version of the continuity equation. Since in the left hand side of Eq.~(\ref{CE1}) we have the time derivative of the on-site number of particles, the right hand side should be identified with a sum of currents flowing along links attached to the site. Indeed introducing a link current from site $\rv$ to site $\rv'$ as follows
\begin{equation} \label{CR1}
J(\rv,\rv')=2 \Im [T_{\rv,\rv'}\: \rho(\rv,\rv';t)],
\end{equation}
we can rewrite the continuity equation \eqref{CE1} in more familiar way
\begin{equation}  \label{CE2}
\dot n(\rv;t)= - \sum_{\rv'} J(\rv,\rv').
\end{equation}
This equation shows that the decrease rate of the density on each site equals to the sum of all outgoing currents. Equation \eqref{CE2} can be also viewed as an integral of the usual differential continuity equation over a small volume element surrounding the site $\rv$.

Now we introduce another equation of primary importance for the lattice TDDFT. This is a lattice analog of a divergence of the local force balance equation. It can be derived by differentiating the continuity equation \eqref{CE1} with respect to time, and using Eq.~(\ref{SE1}) to transform the derivative of the right hand side. After straightforward calculations the lattice force balance equation reduces to the following form
\begin{equation}
 \label{LF1}
\ddot n(\rv;t)=2\Re\sum_{\rv'} T_{\rv,\rv'}\rho(\rv,\rv';t)[v(\rv';t)-v(\rv;t)]-q(\rv;t).
\end{equation}
Here $q(\rv;t)$ stands for a ''lattice divergence'' of an internal stress force, 
\begin{eqnarray} \label{ST1}
q(\rv;t)&=&2\Re\sum_{\rv',\rv''} T_{\rv,\rv'}\Big\{\rho_2(\rv,\rv'',\rv';t)(w_{\rv , \rv''}-w_{\rv',\rv''})\nonumber\\
&+& [T_{\rv',\rv''}\rho(\rv,\rv'';t) - T_{\rv,\rv''}\rho(\rv',\rv'';t)]\Big\},
\end{eqnarray}
where $\rho_2(\rv,\rv'',\rv';t)$ in the right hand side is the two body density matrix,
\begin{eqnarray} \label{rho_2}
\rho_2(\rv,\rv'',\rv';t)= &N&(N-1)\sum_{\rv_3,...,\rv_N} \psi^*(\rv,\rv'',...,\rv_N;t) \nonumber\\
&\times & \psi(\rv',\rv'',...,\rv_N;t).
\end{eqnarray}
A special role of Eq.~\eqref{LF1} for TDDFT follows from the fact that it explicitly relates the potential $v(\rv;t)$ to the density $n(\rv;t)$ and the instantaneous many-body state $\psi(t)$. 

The force balance equation (\ref{LF1}) is the the main result of the present section, which will be used in the next section to analyze the existence of the lattice TDDFT.

\section{TDDFT on a lattice}

The whole concept of TDDFT is based on the existence a one-to-one map between the time dependent density and the external potential. In this section we will prove the mapping and the $v$-representability theorems for the lattice TDDFT by adopting ideas recently proposed for the lattice version of TDCDFT \cite{TokatlyTDCDFT2011}.

\subsection{Statement of the problem and the basic existence theorem}

In general we follow the NLSE approach to TDDFT-type theories \cite{Tokatly2007,Maitra2010,Tokatly2011,TokatlyTDCDFT2011}. Let us first review the basics of this formalism in a more familiar continuum case and then discuss its modifications for many-body dynamics on a lattice. 

In a continuum system the starting point is the usual $N$-particle Schr\"odinger equation  
\begin{equation} \label{SE2}
i \partial _t |\psi (t) \rangle = [ \hat T + \hat W + \hat v (t) ] |\psi (t) \rangle,
\end{equation}
where $\hat T$, $\hat W$ and $\hat v(t)$ are the kinetic energy operator, the interaction  Hamiltonian and the external potential respectively. For a given initial state $|\psi_0 \rangle$ the solution of Eq.~(\ref{SE2}) determines the wave function $|\psi (t) \rangle$ and the density $n(\rv;t)=\langle\psi(t)|\hat{n}(\rv)|\psi (t) \rangle$ as unique functionals of the external potential, i.~e. 
$|\psi [v](t)\rangle$ and $n[v](t)$. Hence within the standard standard statement of the problem in quantum mechanics the linear Schr\"odinger equation defines a unique ``direct map`` from the external potential and the initial state to the time-dependent wave function and the density: $\{v(t),|\psi_0 \rangle\}\mapsto\{n(t),|\psi (t) \rangle\}$. TDDFT assumes that in the above map the potential $v(t)$ and the density $n(t)$ can be interchanged. In other words, TDDFT relies on the existence of an ''inverse map`` from a time-dependent density and the initial state to the time-dependent wave function and a potential that produces the prescribed density: $\{n(t),|\psi_0 \rangle\}\mapsto\{v(t),|\psi (t) \rangle\}$. Constructively the inverse mapping can be defined as follows. Given the initial state $|\psi_0 \rangle$ and the density $n(\rv;t)$ one finds a time-dependent wave function by solving the Schr\"odinger equation (\ref{SE2}), where the potential $v(\rv;t)$ is not fixed externally, but determined self-consistently from the force balance equation
\begin{equation}
\label{force-balance1}
\dfrac{1}{m} \nabla \cdot [n(\rv;t) \nabla v(\rv;t)]=\ddot n(\rv;t)+ q[\psi(t)](\rv).
\end{equation}
In this equation the quantity $q[\psi(t)](\rv)$ (the divergence of the stress force) is the following instantaneous bilinear functional 
of $|\psi (t)\rangle$
\begin{equation}
q[\psi(t)] (\rv)= \nabla\langle \psi (t)|[ \hat{\jv}(\rv),\hat T +\hat W] |\psi (t) \rangle
\end{equation}  
where $\hat{\jv}(\rv)$ is the usual operator of the current density. Apparently the initial state should be consistent with behavior of the density around initial time $t_0$ to satisfy the following relations
\begin{eqnarray}
 \label{n_0}
 \langle\psi_0|\hat{n}(\rv)|\psi_0\rangle &=& n(\rv;t_0) \\
 \label{dotn_0}
 -\nabla\langle\psi_0|\hat{\jv}(\rv)|\psi_0\rangle &=& \dot n(\rv;t_0)
\end{eqnarray}

In this framework the proof of existence of TDDFT reduces to proving the uniqueness and existence of solutions to the nonlinear many-body problem of Eqs.~(\ref{SE2}), (\ref{force-balance1}) supplemented with an initial condition which satisfies the consistency relations of Eqs.~(\ref{n_0}) and (\ref{dotn_0}). 

Strategically the solution of the outlined nonlinear problem contains two major steps: (i) inverting the Sturm-Liouville operator in the left hand side of Eq.~(\ref{force-balance1}) to find the potential as functional of a given density and the instantaneous state, $v[n(t),|\psi(t)\rangle](\rv)$; and (ii) inserting this potential into Eq.~(\ref{SE2}) and solving the resulting NLSE. In a continuum proving the corresponding existence theorems for either step is a highly nontrivial and currently unsolved problem, although a significant progress has been made recently \cite{RugPenBau2009,RugLee2011,Ruggenthaleretal2012}. Below we reformulate the problems (i)-(ii) for the lattice many-body theory  and show that in this case a rigorous proof of existence can be given.

On a lattice the construction of the inverse map $\{n(t),|\psi_0 \rangle\}\mapsto\{v(t),|\psi (t) \rangle\}$ consists of solving the discrete Schr\"odinger equation (\ref{SE1}), where the potential $v(\rv;t)$ is determined self-consistently from the lattice force balance equation of Eq.~(\ref{LF1}) which we rewrite as follows
\begin{equation}
 \label{LF2}
 \sum_{\rv'}k_{\rv,\rv'}[\psi(t)]v(\rv;t) = \ddot n(\rv;t)+ q[\psi(t)](\rv).
\end{equation}
Here the functional $q[\psi(t)](\rv)$ is defined by Eq.~(\ref{ST1}) and we introduced the following notation
\begin{equation} \label{K1}
k_{\rv,\rv'}(\psi)=2\Re\left[T_{\rv,\rv'}\rho(\rv,\rv')-\delta_{\rv,\rv'}\sum_{\rv''}T_{\rv,\rv''}\rho(\rv,\rv'')\right].
\end{equation}
The initial condition for the nonlinear problem of Eqs.~(\ref{SE1}) and (\ref{LF2}) should satisfy the consistency conditions, which follow from the definition of the density Eq.~(\ref{D1}) and the continuity equation (\ref{CE1})
\begin{eqnarray}
 \label{n_0-lat}
 N\sum_{\rv_2,...,\rv_N} \vert \psi_0(\rv,\rv_2,...,\rv_N) \vert ^2 &=& n(\rv;t_0),\\
 \label{dotn_0-lat}
 -2\Im\sum_{\rv'}T_{\rv,\rv'}\rho_0(\rv,\rv') &=& \dot n(\rv;t_0).
\end{eqnarray}

Equations (\ref{SE1}), (\ref{LF2}), (\ref{n_0-lat}), and (\ref{dotn_0-lat}) are the lattice analogs of Eqs.~(\ref{SE2}), (\ref{force-balance1}), (\ref{n_0}), and (\ref{dotn_0}). A dramatic simplification of the lattice theory comes from the fact that both the Hilbert space $\mathcal{H}$ and the space $\mathcal{V}$ of lattice-valued potentials become finite dimensional, with the dimensions $M^N$ and $M$, respectively. In particular the lattice $N$-body Schr\"odinger equation (\ref{SE1}) corresponds to a system of $M^N$ ODE, while the force balance equation (\ref{LF2}) turns out to be a system of $M$ algebraic equations. In fact, Eq.~(\ref{LF2}) can be conveniently rewritten in a matrix form
\begin{equation}\label{LF3}
\hat{K}(\psi)V= S(\ddot n,\psi),
\end{equation}
where $\hat{K}$ is a real symmetric $M\times M$ matrix with elements $k_{\rv,\rv'}$ of Eq.~(\ref{K1}), and $V$ and $S$ are $M$-dimensional vectors with components  
\begin{equation}\label{VandS}
v_{\rv}=v(\rv) \quad {\rm and}\quad s_{\rv}(\ddot n,\psi)=\ddot n(\rv)+q[\psi](\rv),
\end{equation}
respectively. The $\hat{K}$-matrix in Eq.~(\ref{LF3}) is a lattice analog of the Sturm-Liouville operator $m^{-1}\nabla n\nabla$ in Eq.~(\ref{force-balance1}). Hence on a lattice the step (i) in solving our nonlinear many-body problem reduces to a simple matrix inversion, which can be performed provided the matrix $\hat{K}$ is nondegenerate. At this point it is worth noting that because of the gauge invariance $\hat{K}$ matrix of Eq.~(\ref{K1}) always has at least one zero eigenvalue that corresponds to a space-constant eigenvector. Therefore if $\mathcal{V}$ is the $M$-dimensional space of lattice potentials $v(\rv)$, then the invertibility/nondegeneracy of $\hat{K}$ should always refer to the invertibility in an $M-1$-dimensional subspace of $\mathcal{V}$, which is orthogonal to a constant vector $v_C(\rv)=C$. In more physical terms this means that the force balance equation (\ref{force-balance1}) determines the self-consistent potential $v[n,\psi](\rv)$ only up to an arbitrary constant. 

Now we are in a position to formulate and to prove the basic existence and uniqueness theorem on the lattice TDDFT. All statements of the Theorem~1 below refer to the lattice $N$-body problem defined in Sec.~II.

{\it Theorem 1. (existence of the lattice TDDFT)} --- Assume that a given time-dependent density $n(\rv;t)$ is nonnegative on each lattice site, sums up to the number of particles $N$, and has a continuous second time derivative $\ddot n(\rv;t)$. Let $\Omega$ be a subset of the $N$-particle Hilbert space $\mathcal{H}$ where the matrix $\hat{K}(\psi)$ of Eq.~(\ref{K1}) has only one zero eigenvalue corresponding to a space-constant vector. If the initial state $\psi_0\in\Omega$, and at time $t_0$ the consistency conditions of Eqs.~(\ref{n_0-lat}) and (\ref{dotn_0-lat}) are fulfilled, then

\noindent
(i) There is a time interval around $t_0$ where the nonlinear many-body problem of Eqs.~(\ref{SE1}), (\ref{LF2}) has a unique solution that defines the wave function $\psi(t)$ and the potential $v(t)$ as unique functionals of the density $n(t)$ and the initial state $\psi_0$;

\noindent
(ii) The solution of item (i) is not global in time if and only if at some maximal existence time $t^*>t_0$ the boundary of $\Omega$ is reached. 

{\it Proof} --- By the condition of the theorem $\psi_0$ belongs to $\Omega$ where $\hat{K}(\psi)$ has only one trivial zero eigenvalue. Hence there is a neighborhood of $\psi_0$, such that for all $\psi$'s from this neighborhood the matrix $\hat{K}(\psi)$ can be inverted (in the $M-1$-dimensional subspace of $\mathcal{V}$, orthogonal to a constant). In other words, we can solve the force balance equation (\ref{LF3}) as $V= \hat{K}^{-1} S$ and express (up to a constant) the on-site potential in terms of the instantaneous wave function and the density  
\begin{equation} \label{V1}
v[n,\psi](\rv)=\sum_{\rv'}\hat{K}^{-1}_{\rv,\rv'}(\psi)s_{\rv}(\ddot n,\psi). 
\end{equation}  
Substituting this potential into Eq.~(\ref{SE1}) we obtain the following NLSE,
\begin{eqnarray} \label{NLSE1}
i\partial_t \psi(\rv_1,...,\rv_N;t)&=&\sum_{j=1}^N \sum_{\rv} \hat{K}^{-1}_{\rv_j,\rv}s_{\rv} 
\psi(\rv_1,...,\rv_N;t)\nonumber\\
&-&\sum_{i=1}^N\sum_{\xv_i} T_{\rv_i ,\xv_i} \psi(...,\xv_i,...;t)\nonumber\\
&+&\sum_{j > i} w_{\rv_i , \rv_j}  \psi(\rv_1,...,\rv_N;t).
\end{eqnarray}
Equation \eqref{NLSE1} supplemented with the initial condition of Eq.~\eqref{IW} constitutes a universal nonlinear many-body problem, which determines the wave function in terms of the density. Formally it corresponds to a Cauchy problem for a system of $N_{\mathcal{H}}=M^N$ ODE of the following structural form
\begin{equation} \label{NLODE}
\dot \psi= F(\psi,t), \qquad \psi(t_0)=\psi_0,
\end{equation}
where $\psi$ is a $N_\mathcal{H}$-dimensional vector living in the Hilbert space $\mathcal{H}$ and the right hand side is a nonlinear function of the $\psi$'s components.

The nonlinearity of $F(\psi,t)$ in Eq.~(\ref{NLSE1}) comes from the dependence of the potential $v[n,\psi](\rv)$ in Eq.~(\ref{V1}) on the wave function, which, in turn, is determined by the functions $\hat{K}^{-1}(\psi)$ and $S(\ddot n,\psi)$. Both $k_{\rv,\rv'}(\psi)$ of Eq.~(\ref{K1}), and $s_{\rv}(\ddot n,\psi)$ of Eq.~(\ref{VandS}) are linear in the density matrices, and thus bilinear in $\psi$ forms. Therefore the potential $v[n,\psi](\rv)$ and, as a consequence, the whole right hand side $F(\psi,t)$ in Eq.~(\ref{NLODE}) are rational functions of the components of the wave function. Moreover the denominator of these rational functions never turns into zero for all $\psi\in\Omega$, which implies that in $\Omega$ the function $F(\psi,t)$ satisfies a uniform Lipshitz condition. An explicit time dependence of $F(\psi,t)$ is determined by the time dependence of $\ddot n(t)$ that is continuous by the condition of the theorem. Thus we conclude that for all $\psi\in\Omega$ the right hand side $F(\psi,t)$ in Eq.~(\ref{NLODE}) is Lipshitz in $\psi$ and continuous in time. 

After identifying the subset $\Omega$ with the domain of Lipshitz continuity we can directly employ the standard results of the theory of nonlinear ODE. Namely, if the initial state $\psi_0\in\Omega$, the Picard-Lindel\"of theorem \cite{DE} guarantees the existence of a finite interval $t_0 -\delta < t < t_0 +\delta$, with $\delta >0$, where the initial value problem \eqref{NLODE} has a unique solution. This solution defines a unique map $\{n(t),\psi_0\}\mapsto\{v(t),\psi (t)\}$ locally in time, in accordance with the statement (i) of the theorem. 

The extension theorems for nonlinear ODE imply that a local solution can not be extended beyond some maximal existence time $t^*>t_0$ only in two cases: first, at $t\to t^*$ the solution becomes unbounded or, second, at $t\to t^*$ it reaches the boundary of $\Omega$. In our case the solution is guarantied to be normalized and thus bounded. Therefore we are left only with the second possibility, which proves the statement (ii) and completes the proof of the theorem.

\subsection{Discussion and comments on the existence theorem}

\subsubsection{Definition of the $v$-representability subset $\Omega$}

According to the Theorem~1, any sufficiently smooth density $n(\rv;t)$ is $v$-representable, at least locally, if the dynamics starts inside the subset $\Omega$ of the Hilbert space. In general to ensure that a state $\psi$ belongs to $\Omega$ we need to check the invertibility of matrix $\hat{K}(\psi)$, which, though possible in principle, may become difficult in practice, especially for lattices with a large number of sites. Is it possible to formulate simpler, but possibly more restrictive criteria, which would guarantee the validity of the lattice TDDFT. 

One simple necessary condition immediately follows from the form of the matrix elements $k_{\rv,\rv'}$ in Eq.~(\ref{K1}). The matrix $\hat{K}(\psi)$ is nondegenerate only if a lattice state $\psi$ is connected in a sense that any two sites can be connected by a line composed of links with nonzero values of $T_{\rv,\rv'}\Re\rho(\rv,\rv')$. Indeed, for a disconnected state $\hat{K}(\psi)$ takes a block-diagonal form and new zero eigenvalues, corresponding to piecewise constant in space eigenvectors, appear. The number of such zero eigenvalues equals to the number of disconnected regions on a lattice. We emphasize that a purely geometric connectivity of the lattice does not automatically guaranties $v$-representability -- two sites connected geometrically by a nonzero hopping matrix element $T_{\rv,\rv'}$ can be disconnected in the above sense if for a state $\psi$ the quantity $\Re\rho(\rv,\rv')$ vanishes. An explicit example of such a disconnected (one-particle) state on a connected 4-site tight-binding cluster has been recently proposed in Ref.~\onlinecite{Kurth2011} to demonstrate a possible non-$v$-representability in the lattice TDDFT. The authors considered an excited state with nodes on two opposite corners of a square formed by four sites (see Fig.~1 in Ref.~\onlinecite{Kurth2011}). The two nodes effectively separate the system into two disconnected parts. Therefore the matrix $\hat{K}$ acquires an extra zero eigenvalue and the Theorem~1 does not apply if the dynamics starts from such a state. In fact, one can show that this particular state lies precisely at the border of the $v$-representability subset $\Omega$.

Obviously the connectivity of the lattice state is only a necessary, but not a sufficient condition for $\psi$ to be in $\Omega$. The reason is that for a connected state the quantities $T_{\rv,\rv'}\Re\rho(\rv,\rv')$ for different links may have different signs which can be responsible for extra zero eigenvalues of $\hat{K}$. Hence the simplest sufficient condition is a connectivity of the state and positivity (or negativity) of $T_{\rv,\rv'}\Re\rho(\rv,\rv')$ for all lattice links. In other words, a state $\psi\in\Omega$ if its $\hat{K}$ matrix is primitive and does not have a block-diagonal form. This condition is easy to check in practice, but it appears to be quite restrictive. A less restrictive criterion that in many cases can still be checked easily, is a positive (negative) definiteness of $\hat{K}(\psi)$. In Sec.~IV we will show that this is exactly the case for a many-body ground state on a connected lattice. Namely, if $\psi_0$ is a ground state, then $\hat{K}(\psi_0)$ is negative definite and thus $\psi_0\in\Omega$, which implies the existence of the lattice TDDFT for a system evolving from its ground state.  

\subsubsection{Boundness of $\dot n(\rv;t)$ in the lattice TDDFT}

A specific feature of quantum dynamics on a lattice, which narrows a class of $v$-representable densities, is a boundness of the time derivative of the density \cite{Baer2008,LiUll2008,Verdozzietal2011,TokatlyTDCDFT2011}. Since the hopping rate along a given link is fixed to be $T_{\rv,\rv'}$, a link current of Eq.~(\ref{CR1}) can not exceed a certain maximal value, i.~e. $|J(\rv,\rv')|\le |J_{\rm{max}}(\rv,\rv')|$, where 
$|J_{\rm{max}}|$ can be estimated \cite{Verdozzi2008,Verdozzietal2011,TokatlyTDCDFT2011} using the Cauchy-Schwarz inequality 
\begin{equation}\label{Jmax}
|J_{\rm{max}}(\rv,\rv')| = 2| T_{\rv,\rv'}\rho(\rv,\rv')|\leq 2|T_{\rv,\rv'}|\sqrt{n(\rv)n(\rv')}.
\end{equation}      
A physical density should satisfy the continuity equation (\ref{CE2}) which imposes a bound on its time derivative,
\begin{equation}
 \label{dot-n-max}
 |\dot n(\rv)| \le \sum_{\rv'} |J_{\rm{max}}(\rv,\rv')|.
\end{equation}

On the first sight the Theorem~1 does not say anything about the boundness of $\dot n(\rv;t)$. Therefore it is instructive to see how it can be deduced from the conditions of the theorem. First of all we note that if a solution of the universal NLSE exists, then the continuity equation is necessarily satisfied, which can be true only if our given density does not violate the bound of Eq.~(\ref{dot-n-max}). According to the assumptions of the theorem the density should satisfy the consistency conditions, Eqs.~(\ref{n_0-lat}) and (\ref{dotn_0-lat}), and its second time derivative $\ddot n(\rv;t)$ should be continuous in time for all $t>t_0$. By imposing the condition of Eq.~(\ref{dotn_0-lat}) we explicitly require the boundness of $\dot n(\rv;t_0)$ at the initial time $t_0$, while the continuity of $\ddot n(\rv;t)$ ensures that the physical bound of Eq.~(\ref{dot-n-max}) can not be violated immediately. It is also worth noting that the boundness of $\dot n(\rv;t)$ is closely related to the invertibility of $\hat{K}$ matrix or, more precisely, to the connectivity of the instantaneous state $\psi(t)$. Indeed, the link current $J(\rv,\rv')$ of Eq.~(\ref{CR1}) and the off-diagonal element $k_{\rv,\rv'}$ of Eq.~(\ref{K1}) are, respectively, the imaginary and the real parts of the quantity $2T_{\rv,\rv'}\rho(\rv,\rv';t)$. Therefore for any state $\psi$ and $\rv\ne\rv'$ the following identity holds true
\begin{equation}
 \label{J-K-relation}
 |J(\rv,\rv')|^2 + |k_{\rv,\rv'}|^2 = |J_{\rm{max}}(\rv,\rv')|^2.
\end{equation}
Equation (\ref{J-K-relation}) shows that when the current $J(\rv,\rv')$ reaches the maximal value of Eq.~(\ref{Jmax}), $k_{\rv,\rv'}$ turns into zero, which breaks the link between sites $\rv$ and $\rv'$. Hence saturation of the bound in Eq.~(\ref{dot-n-max}) implies breaking all links attached to a site $\rv$. This site becomes disconnected from the rest of the lattice and the $\hat{K}$ matrix acquires an extra zero eigenvalue, indicating that the state $\psi$ is not anymore in the $v$-representability subset $\Omega$. Thus saturation of the bound on $\dot n(\rv)$ at some time $t=t^*$ automatically assumes that at this time the solution hits the boundary of $\Omega$. This behavior is in a clear agreement with the statement (ii) of Theorem~1.

\section{Explicit illustration: One particle on a two-site lattice}

This section is aimed at illustrating the general NLSE scheme using a simple exactly solvable example -- one particle on two sites. In spite of its simplicity this example contains practically all features of the most general $N$-body case, and thus displays all subtle points of the general formulation in a clear and transparent form. Therefore it is advisable to read this section to get better feeling of the formalism presented in Sec.~III. Another explicit example of NLSE for an interacting two-particle system is given in Appendix.

Consider a particle living on a two-site lattice. The state of the system is described by the one-particle wave function 
$\psi_{\rv}(t)$, where the coordinate $\rv$ takes values 1 or 2 corresponding to the two lattice sites.

The dynamics of the system is described by Eq.~\eqref{SE1} where the number of the particles $N=1$ and there is no interaction term in the right hand side. Therefore Eq.~\eqref{SE1} reduces to the following system of two ODEs.
\begin{subequations}
\label{1SE}
\begin{align} 
i \partial_t  \psi_1  = v_1 \psi_1  - T \psi_2 ,\label{1SEa} \\
i \partial_t  \psi_2  = v_2 \psi_2  - T \psi_1 ,\label{1SEb}
\end{align} 
\end{subequations}
where $T$ is a real hopping rate, and $v_1$ and $v_2$ are the time-dependent external potentials on sites 1 and 2, respectively.
The system of Eqs.~\eqref{1SE} determines the components of the wave function $\psi_1$ and $\psi_2$ as functionals of the external potential and the initial state $\psi_{1,2} (0)$ (here we set $t_0=0$).

To find the wave function as a functional of the density we have to construct the proper NLSE, and for this one needs an additional equation which relates the potential to the density and the wave function. In the general framework of Sec.~III the force balance equation of Eq.~\eqref{LF1} [or, equivalently, Eq.~(\ref{LF3})] serves exactly for this purpose. For two sites $\hat{K}$ is a 
$2\times 2$ matrix and therefore Eq.~(\ref{LF3}) takes the form 
\begin{equation} \label{1LF}
 \begin{pmatrix}
  -k_{12} & k_{12} \\
  k_{12} & -k_{12} 
 \end{pmatrix}
 \begin{pmatrix}
v_1\\
v_2
 \end{pmatrix}
 =
 \begin{pmatrix}
 \ddot n_1 + q_1 \\
\ddot n_2 + q_2
 \end{pmatrix}
\end{equation}
where $k_{\rv \rv'}$ and $q_{\rv}$ can easily  be derived from the general definitions of Eqs.~\eqref{K1} and \eqref{ST1}, respectively, 
\begin{eqnarray}
k_{12}&=&2 T \:\Re \rho_{12},\label{k12}\\
q_1=-q_2&=& -2 T ^2 (|\psi_2 |^2 - |\psi_1 |^2).\label{1q}
\end{eqnarray}
The link density matrix $\rho_{12}$, which is in general determined by Eq.~\eqref{rho1}, in the present one-particle case reduces to a simple product $\rho_{12} = \psi_1^*\psi_2$. The $2\times 2$ matrix in right hand side of Eq.~\eqref{1LF} is the $\hat K$ matrix extensively discussed in the previous section. Obviously it always has a zero eigenvalue corresponding to a space-constant potential $v_1=v_2=C$. If there is no other zero eigenvalue we can invert $\hat K$ in a space perpendicular to the constant vector $v_C (\rv)=C$.  Being perpendicular to $v_C (\rv)$ simply means that on-site potentials sum up to zero, which for two sites implies
\begin{equation}\label{v}
v_1=-v_2=v.
\end{equation}
This equation defines a 1-dimensional subspace of $\mathcal{V}$ where $\hat K$ can in principle be inverted. In the present two-site case the $\hat K$ matrix is invertible if $k_{12}\neq 0$. Therefore the $v$-represenatability subset $\Omega$ of the Hilbert space is defined by the following simple condition
\begin{equation} \label{IE1}
\Re [\psi_1^*  \psi_2] \neq 0.
\end{equation}
For all states satisfying the condition Eq.~(\ref{IE1}) we can invert $\hat K$ matrix in Eq.~\eqref{1LF} and find the potential $v=v_1=-v_2$ as a functional of the density and the wave function,
\begin{equation} \label{1v}
 v =- \dfrac{\ddot n_1 - 2 T ^2 (|\psi_2 |^2 - |\psi_1|^2)}{4T \Re \rho_{12}}.
\end{equation}
where we substituted explicit expressions for $k_{12}$ and $q_1$ from Eqs.~\eqref{k12} and (\ref{1q}).

The final NLSE is obtained by inserting the potential of Eq.~\eqref{1v} into the Schr\"odinger equation \eqref{1SE}
 \begin{subequations}
\label{1NLSE}
\begin{eqnarray} 
i \partial_t \psi_1 &=&- \dfrac{\ddot n_1 - 2 T ^2 (|\psi_2 |^2 - |\psi_1 |^2)}{4T \Re[\psi_1^* \psi_2]} \psi_1 - T \psi_2, \label{1NLSEa}\\
i \partial_t \psi_2 &=& \dfrac{\ddot n_1 - 2 T^2 (|\psi_2 |^2 - |\psi_1 |^2)}{4T \Re [\psi_1^* \psi_2]} \psi_2  - T\psi_1.\label{1NLSEb}
\end{eqnarray}
\end{subequations}
This system of equations perfectly illustrates the generic structure of NLSE appearing in the TDDFT context. Firstly, as described in Sec.~III, the nonlinearity is always a rational function with enumerator and denominator being bilinear forms in the components of the wave function $\psi$. For all $\psi\in\Omega$ [i.~e. for $\psi$ satisfying Eq.~(\ref{IE1})] the denominator never turns into zero. Secondly, the explicit time dependence enters NLSE only via the second time derivative of the density $\ddot n_{\bf r}(t)$ that is assumed to be continuous. The above two properties ensure that the right hand side of our NLSE is Lipshitz in $\psi$ and continuous in $t$. By the Picard-Lindel\"of theorem this guarantees the existence of a unique solution to Eqs.~(\ref{1NLSE}) for any initial state $\psi(0)$ from $\Omega$. 

However, this is not yet the whole story. Since the density enters the equations only via $\ddot n_{\bf r}(t)$, our unique solution to NLSE, in general, will reproduce correctly only the second time derivative of the prescribed density. The whole externally given density $n_{\bf r}(t)$ is recovered from NLSE if the dynamics starts from a special manifold of the ``density-consistent'' initial states which are defined by the consistency conditions of Eqs.~\eqref{n_0-lat} and \eqref{dotn_0-lat}.

To proceed further with our example we represent the wave function in the polar form 
\begin{equation}
 \label{psi-polar}
 \psi_{1}(t)=|\psi_{1}(t)| e^{i\varphi(t)/2}, \: \psi_{2}(t)=|\psi_{2}(t)| e^{-i\varphi(t)/2},
\end{equation}
and substitute it into the consistency conditions. As a result Eqs.~\eqref{n_0-lat} and \eqref{dotn_0-lat}, respectively, simplify as follows
\begin{eqnarray}
\label{psi0}
|\psi_{\bf r}(0)| &=& \sqrt{n_{\bf r} (0)}, \quad {\bf r}=\{1,2\},\\
\label{phi0}
\dot n_1 (0) &=& 2T \sqrt{n_1(0) n_2(0) } \sin \varphi(0),
\end{eqnarray}
where $n_{\rv} (0)$ is the (prescribed) initial density, and $\dot n_{\bf r} (0)$ its initial time derivative. The first condition, Eq. \eqref{psi0}, uniquely determines the modulus of the allowed initial states in terms of the initial density. Finding the ``density-consistent`` initial phases is a bit more tricky as the right hand of Eq.~\eqref{phi0} is not a single valued function. Equation~\eqref{phi0} has two solutions which can be written in the following form
\begin{eqnarray}
\varphi^{(+)}(0) &=&\arcsin \left(\dfrac{\dot n_1(0)}{2 T \sqrt{n_1(0)n_2(0)} }\right) \equiv \phi_0,\label{phi01}\\
\varphi^{(-)}(0) &=&\pi - \phi_0, \label{phi02}
\end{eqnarray}
where  $\arcsin$ stands for the principal value of the inverse of $\sin$. In other words, $\phi_0$ defined after Eq.~(\ref{phi01}) is a solution to Eq.~\eqref{phi0} in the interval $[-\frac{\pi}{2},\frac{\pi}{2}]$. The existence of two solution to the consistency conditions means that for a given density in our simple system the manifold of density-consistent initial states consists of the two following wave functions
\begin{eqnarray}
\label{psi0+}
\psi_{1}^{(+)}(0)=\sqrt{n_{1} (0)} e^{i\frac{\phi_0}{2}}, \: \psi_{2}^{(+)}(0)=\sqrt{n_{2} (0)} e^{-i\frac{\phi_0}{2}} &\:&\\
\label{psi0-}
\psi_{1}^{(-)}(0)=\sqrt{n_{1} (0)} e^{-i\frac{\phi_0}{2}}, \: \psi_{2}^{(-)}(0)=-\sqrt{n_{2} (0)} e^{i\frac{\phi_0}{2}} &\:&
\end{eqnarray}
where we disregarded an irrelevant common phase factor. By substituting Eqs.~(\ref{psi0+}) and (\ref{psi0-}) into the condition of Eq.~(\ref{IE1}) we find that the initial states $\psi^{(\pm)}(0)\in\Omega$ provided $\phi_0\ne\pm\pi/2$. Obviously this puts a restriction on the initial values of $n$ and $\dot n$. Equation~(\ref{phi01}) tells us that the condition $\phi_0\ne\pm\pi/2$ actually ensures that $\dot n(0)$ is properly bounded [see Eqs.~(\ref{Jmax}) and (\ref{dot-n-max})]. 

It is interesting to note that if $\dot n(0)=0$, than $\phi_0=0$ and the density-consistent initial states Eqs.~(\ref{psi0+}) and (\ref{psi0-}) can be viewed as the ground (symmetric) and the excited (antisymmetric) states of a dimer in the presence of some static potential. 

Now we can solve NLSE of Eq.~(\ref{1NLSE}) starting from one of the allowed initial states. The solution should return the wave function and the potential as unique functionals of the given density $n_{\bf r}(t)$. Inserting the polar representation Eq.~(\ref{psi-polar}) into Eq.~(\ref{1NLSE}) we observe that the following form of $\psi_{\bf r}(t)$,
\begin{equation}
 \label{NLSEsolution}
 \psi_{1}(t)=\sqrt{n_{1}(t)} e^{i{\varphi(t)}/{2}}, \: \psi_{2}(t)=\sqrt{n_{2}(t)} e^{-i{\varphi(t)}/{2}},
\end{equation}
solves NLSE if the time-dependent phase $\varphi(t)$ satisfies the equation
\begin{equation}
 \label{phi-dynamic}
 \dot n_1 (t) = 2T \sqrt{n_1(t) n_2(t) } \sin \varphi(t).
\end{equation}
For each initial state from the set of Eqs.~(\ref{psi0+}) and (\ref{psi0-}) this equation has a unique solution provided, the condition
$k_{12}(\psi)\neq 0$ is fulfilled.

Assume that we started from the state $\psi_{1}^{(+)}(0)$, Eq.~(\ref{psi0+}). Then the solution to Eq.~(\ref{phi-dynamic}) reads
\begin{equation} \label{1phi}
 \varphi =\arcsin \left(\frac{\dot n_1}{2 T \sqrt{n_1n_2}}\right).
\end{equation}
This equation together with Eq.~\eqref{NLSEsolution} gives the wave function as a functional of the density. As long as this solution exists, the element 
$k_{12}=2T\Re \rho_{12}$ of the $\hat{K}$ matrix stays positive
\begin{equation}
\label{k12+}
k_{12}^{(+)} = 2T \sqrt{n_1 n_2} \cos \varphi = \sqrt{4 T^2 n_1n_2-\dot n^2_1}
\end{equation}

To find the potential $v(t)$ as a functional of the density we insert Eqs.~(\ref{NLSEsolution}) and (\ref{k12+}) into Eq. \eqref{1v}. The results takes the following form
\begin{eqnarray} \label{1v+}
 v^{(+)}[n]= - \frac{\ddot n_1 - 2 T ^2 (n_1 - n_2)}{\sqrt{4 T^2 n_1n_2-\dot n^2_1}}.
\end{eqnarray}
This functional reproduces with the result obtained in Ref.~\onlinecite{LiUll2008}. In addition there is another solution that corresponds to another density-consistent initial state.

If we start from the second initial state, $\psi_{1}^{(-)}(0)$ of Eq.~(\ref{psi0-}), we should take the second solution of Eq.~(\ref{phi-dynamic}) for the phase, namely
\begin{equation} \label{phi}
 \varphi = \pi - \arcsin \left(\frac{\dot n_1}{2 T \sqrt{n_1n_2} }\right).
\end{equation}
In this case $k_{12}$ changes a sign, 
\begin{equation} 
\label{k12-}
k_{12}^{(-)}= -\sqrt{4 T^2 n_1n_2-\dot n^2_1},
\end{equation}
which implies that the sign of the potential $v$ is also reversed 
\begin{eqnarray}  \label{1v-}
 v^{(-)}[n]= \dfrac{\ddot n_1 - 2 T ^2 (n_1 - n_2)}{\sqrt{4 T^2 n_1n_2-\dot n^2_1}}.
\end{eqnarray}
Thus, for different initial conditions the NLSE machinery produces unambiguously different functionals $v[n]$ and $\psi[n]$. This nicely displays the initial state dependence in TDDFT \cite{MaiBurWoo2002,EllMai2012}, although in the present case the dependence is very simple.

For each density-consistent initial state the unique solution to NLSE of Eq.~(\ref{1NLSE}) exists as long as the boundary of the subset $\Omega$ is not reached. This happens if $k_{12}(\psi)=0$, i.~e. when the expression under the square root in Eqs.~(\ref{k12+}) or (\ref{k12-}) turns into zero. In agreement with the general discussion in Sec.~III, at this point the bound on the time derivative of the density,
\begin{equation}
|\dot n_1 | < 2 T \sqrt{n_1n_2},
\end{equation} 
is saturated.

In our simple model we can also visualize and completely characterize the geometry of the $v$-representability subset $\Omega$ in the Hilbert space $\mathcal{H}$. Since we have an effective 2-level system the projective Hilbert space can be represented by a 2-sphere, known as a Bloch sphere. Specifically, after taking out a common phase factor, each normalized state from $\mathcal{H}$ is mapped to a point on a 2-sphere in $\mathbb R^3$ (see Fig.~1) by parametrizing the wave function as follows 
\begin{equation} 
\label{psiS}
|\psi\rangle = \cos {\theta/2} e^{i \varphi/2} \; |1 \rangle +\sin{\theta/2}  \; e^{-i \varphi/2} |2\rangle,
\end{equation}
where $|1 \rangle$ and $|2 \rangle$ are the orthogonal states corresponding to the particle residing on sites 1 and 2, respectively. In this mapping the moduli of the on-site amplitudes are represented by the azimuthal angle $\theta$, while the phase difference $\varphi$ corresponds to the polar angle in spherical coordinates. As we can see from Fig.~1 the line $k_{12}(\psi)=0$ divides the projective Hilbert space into two, left and right, hemispheres with $k_{12}>0$ and $k_{12}<0$. The two hemispheres represent two disconnected parts of the $v$-representability subset $\Omega$, separated by the boundary line $k_{12}=0$. The boundary line contains all states for which the statements of the Theorem~1 do not hold. Starting from any point on those hemispheres, i.~e. from a state $\psi\in\Omega$, we uniquely recover the time evolution of the system with a given density by solving NLSE of Eq. \eqref{1NLSE}. As long as the trajectory stays within the original hemisphere and does not touch the boundary, the one-to-one density-to-potential map exists with the functional $v[n]$ given by Eq.~(\ref{1v+}) or by Eq.~(\ref{1v-}), depending on the hemisphere. Whether it is possible to construct a unique and universal analytic continuation for crossing the boundary and covering the whole subset $\Omega$ is an interesting question, which can not be answered at the level of Theorem~1.
\begin{figure} 
\includegraphics[scale=0.4]{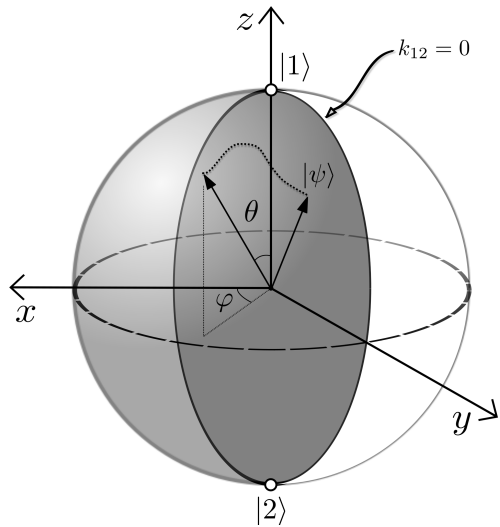}
\label{fig:Bloch}
\caption{Each normalized state in the Hilbert space $\mathcal H$ maps to a point on the Bloch sphere. The north $|1 \rangle$ and the south $|2\rangle$ poles correspond to the particle on sites 1 and 2. The line $k_{12}=0$ divides the sphere into two (left and right) hemispheres corresponding to two disconnected parts of the $v$-representability subset $\Omega$.}
\end{figure}

\section{Time-dependent $v$-representability for a system evolving from the ground state}

In this section we return to the most general case, and show that the ground state of a lattice $N$-particle system always belongs to the $v$-representability subset $\Omega$. This implies that the lattice TDDFT is guaranteed to exist if the dynamics starts from the ground state.

Assume that $\psi_k=|k\rangle$ form a complete set of eigenstates for a lattice many-body Hamiltonian describing $N$-particle system in the presence of a static scalar potential $v_0(\rv)$. Let $\psi_0=|0\rangle$ be the ground state. We are going to prove that the matrix $\hat{K}(\psi_0)$ evaluated at the ground state is strictly negative definite in the subspace of potentials that are orthogonal to a space-constant vector $V_{C}$, i.~e.,
\begin{equation}
 \label{K-negativity1}
 V^{T}\hat{K}(\psi_0)V \equiv \sum_{\rv,\rv'}v({\rv})k_{\rv,\rv'}v({\rv'}) < 0,
\end{equation}
for all $M$-dimensional vectors $V=\{v({\rv})\}$ which satisfy the following orthogonality relation
\begin{equation}
 \label{v-ortho}
 V^{T}V_{C} = C\sum_{\rv}v({\rv})=0,
\end{equation}
where $V^T$ stands for a transposed vector.

Our starting point is the $f$-sum rule (see, e.~g., Ref.~\onlinecite{Vignale2005}) for the density-density response function 
$\chi_{\rv,\rv'}(\omega)=\langle\langle\hat{n}_{\rv};\hat{n}_{\rv'}\rangle\rangle_{\omega}$:
\begin{equation} \label{f-sumrule}
- \dfrac{2}{\pi} \int_0 ^{\infty}\omega \Im\chi_{\rv,\rv'}(\omega)d \omega = 
i \langle 0|[\dot{\hat{n}}_{\rv},\hat{n}_{\rv'}]|0\rangle .
\end{equation} 
To calculate the commutator in the right hand side of Eq.~(\ref{f-sumrule}) we switch to the second quantized representation and write the equation of motion for the density operator $\hat{n}_{\rv}=\hat{a}_{\rv}^\dagger\hat{a}_{\rv}$,   
\begin{equation}
\label{continuity2}
\dfrac{d \hat{n}_{\rv}}{d t}=i \sum_{\rv'} (T_{\rv,\rv'} \hat{a}_{\rv} ^\dagger \hat{a}_{\rv'}-T_{\rv',\rv} \hat{a}_{\rv'} ^\dagger \hat{a}_{\rv}),
\end{equation}
where ${a}_{\rv}$ and $\hat{a}_{\rv}^\dagger$ are the on-site annihilation and creation operators. Equation~(\ref{continuity2}) is nothing but the operator form of the continuity equation. Using Eq.~(\ref{continuity2}) one can easily calculate the commutator entering the right hand side of Eq.~\eqref{f-sumrule},
\begin{equation}
i[\dot {\hat{n}} _{\rv},\hat{n}_{\rv'}] =- T_{\rv,\rv'}\hat{a}_{\rv} ^\dagger \hat{a}_{\rv'} + 
\delta_{\rv\rv'} \sum_{\rv''} T_{\rv,\rv''}\hat{a}_{\rv}^\dagger \hat{a}_{\rv''} + h.c.
\end{equation}
Taking the ground state expectation value of this equation and comparing the result with Eq.~(\ref{K1}) we find
\begin{equation} \label{K}
i \langle 0|[\dot {\hat{n}} _{\rv},\hat{n}_{\rv'}]|0\rangle = -k_{\rv,\rv'}.
\end{equation}
Therefore the right hand side of the lattice $f$-sum rule is identified with the $\hat{K}$ matrix entering the definition of the $v$-representability subset \cite{Note1}.

On the other hand, for the imaginary part of density response function we have the following spectral representation \cite{Vignale2005}
\begin{equation} \label{momentg}
\int\limits_0 ^{\infty}\omega \Im\chi_{\rv,\rv'}(\omega)d \omega =  - 2\pi \Re \sum_{k} \omega_{k0}\langle 0| \hat{n}_{\rv}| k \rangle \langle k| \hat{n}_{\rv'}|0\rangle,
\end{equation}
where  $\omega_{k0}=E_k-E_0$ are excitation energies of the system. Substitution of Eqs.~(\ref{K}) and (\ref{momentg}) into Eq.~(\ref{f-sumrule}) leads to the spectral representation for the elements of $\hat{K}$ matrix
\begin{equation}\label{Kground}
k_{\rv,\rv'}= -4\Re\sum_{k} \omega_{k0}
\langle 0|\hat{n}_{\rv}|k\rangle \langle k|\hat{n}_{\rv'}|0\rangle.
\end{equation}
Finally, inserting $k_{\rv,\rv'}$ of Eq.~(\ref{Kground}) into the left hand side of Eq.~(\ref{K-negativity1}), we arrive at the following remarkable result
\begin{eqnarray}
\nonumber
V^{T}\hat{K}(\psi_0)V &=&  -4\sum_{k}\omega_{k0}{\Big |}\sum_{\rv} v({\rv})\langle 0|\hat{n}_{\rv}|k\rangle{\Big |}^2\\
\label{K-negativity2}
 &=&-4\sum_{k}\omega_{k0}|\langle 0|\hat{v}|k\rangle|^2 \le 0,
\end{eqnarray}
where $\hat{v}$ is a many-body operator corresponding to the potential $v({\rv})$,
\begin{equation}
 \label{v-operator1}
 \hat{v} = \sum_{\rv} v({\rv})\hat{n}_{\rv}.
\end{equation}

Let us show that the equality in Eq.~(\ref{K-negativity2}) holds only for a space-constant potential $v_C(\rv)=C$. Since each term in in the sum in Eq.~(\ref{K-negativity2}) is non-negative, the result of summation is zero if and only if
\begin{equation}
 \label{v_0k}
 \langle 0|\hat{v}|k\rangle=0, \quad \text{for all $k\ne 0$}.
\end{equation}
Physically the right hand side of Eq.~(\ref{K-negativity2}) is proportional to the energy absorbed by a system after a small amplitude pulse of the form $v(\rv;t)=v(\rv)\delta(t)$. Then the condition Eq.~(\ref{v_0k}) simply states that nothing is absorbed only if the potential $v(\rv)$ does not couple the ground state to any of the excited states. 

Assume that Eq.~(\ref{v_0k}) is fulfilled and expand a vector $\hat{v}|0\rangle$ in a complete set of states $\{|k\rangle\}$
\begin{equation}
 \label{VPsi0}
 \hat{v}|0\rangle = \sum_k |k\rangle\langle k|\hat{v}|0\rangle = |0\rangle\langle 0|\hat{v}|0\rangle \equiv \lambda|0\rangle.
\end{equation}
Therefore the condition of Eq.~(\ref{v_0k}) implies that the ground state $|0\rangle$ is an eigenfunction of the operator $\hat{v}$. Since $\hat{v}$ corresponds to a local multiplicative one-particle potential this can happen only if the potential is a constant. For clarity we write Eq.~(\ref{VPsi0}) in the coordinate representation
\begin{equation}
 \label{VPsi02}
 \sum_{j=1}^N v(\rv_j)\psi_0(\rv_1,...,\rv_N) = \lambda \psi_0(\rv_1,...,\rv_N).
\end{equation}
Obviously this equation can be fulfilled only if the function $v(\rv)$ takes the same value $\lambda/N$ on all lattice sites, which corresponds to a space-constant potential. A notable exception is a geometrically disconnected lattice consisting of a several pieces that can not be connected by a path composed of links with nonzero $T_{\rv,\rv'}$. In this case the arguments of the wave function $\psi_0(\rv_1,...,\rv_N)$ form ``disconnected groups'' of coordinates corresponding to particles residing in disconnected parts of the system. The coordinates of different disconnected groups take values in ``non-overlapping'' parts of the lattice. Since the number of particles in each part (number of coordinates in each group) is fixed, Eq.~(\ref{VPsi02}) can also be satisfied with a piecewise constant potential.

Therefore we arrive at the following conclusion: for a connected lattice Eq.~(\ref{v_0k}) is fulfilled, and the inequality in Eq.~(\ref{K-negativity2}) is saturated only for a constant in space potential. For all potentials which are orthogonal to a constant in a sense of Eq.~(\ref{v-ortho}) the strict inequality in Eq.~(\ref{K-negativity2}) takes place. This means that matrix 
$\hat{K}(\psi_0)$ is negative definite and thus invertible in the $M-1$-dimensional subspace of $\mathcal{V}$ orthogonal to a constant vector $V_C$. In other words, the ground state of $N$-particle system on a connected lattice does belong to the $v$-representability subset $\Omega$. This result combined with the general existence theorem of Sec.~III proves the following particular version of the time-dependent $v$-representability theorem.

{\it Theorem 2.} --- Let the initial state $\psi_0$ for a time-dependent many-body problem on a connected lattice corresponds to a ground state in the presence of some scalar potential $v_0(\rv)$. Then any density $n(\rv;t)$ which satisfies the consistency conditions of Eqs.~(\ref{n_0-lat}) and (\ref{dotn_0-lat}) and has a continuous second time derivative is locally $v$-representable.

It is worth noting that the above ground state based argumentation can be straightforwardly extended to a thermal equilibrium state. In this case Eq.~(\ref{K-negativity2}) takes the form
\begin{equation}
\label{Kbeta}
V^{T}\hat{K}^{\beta}V = 4\sum_{k>l}\omega_{kl}\frac{e^{-\beta E_k}-e^{-\beta E_l}}{\mathcal{Z}}
|\langle l|\hat{v}|k\rangle|^2,
\end{equation}
where $\beta$ is inverse temperature, $\hat{K}^{\beta}$ is the $\hat{K}$ matrix evaluated for the thermal equilibrium state, $\mathcal{Z}$ is partition function, and $\omega_{kl}=E_k-E_l$. By the same token the form defined by Eq.~(\ref{Kbeta}) is strictly negative for all potentials orthogonal to a constant vector. Therefore the Theorem~2 should also apply to the ensemble version of TDDFT based on the von~Neuman equation for the $N$-body density matrix. Of course in this case we need to prove an ensemble extension of Theorem~1, but currently this also looks relatively straightforward.

\section{Conclusion}

In conclusion we presented a rigorous formulation of the lattice TDDFT. On the technical side we mapped the problem of existence of TDDFT to the Cauchy problem for a special NLSE and proved the corresponding existence and uniqueness theorem using standard results from the theory of nonlinear ODE. As one could expect from the previous works \cite{Kurth2011,LiUll2008,Baer2008,Verdozzi2008,Verdozzietal2011} the lattice TDDFT is not unconditionally valid. Similarly to the lattice TDCDFT, apart from a certain quite mild restrictions on the time dependence of the density, there are also limitations on possible many-body initial states. Namely, allowed initial states form a special $v$-representability subset $\Omega$ of the Hilbert space $\mathcal{H}$. Most physical states should be in $\Omega$. In particular we demonstrated that this is true for any ground state on a connected lattice. Therefore in practice for most non-exotic physical situations the lattice TDDFT (and TDCDFT) should be valid. In this context it is interesting to note a recently observed example of a violation of the non-interacting $v$-representability for a strongly biased interacting resonant level model \cite{SchDziSch2012}.

Our proof of the basic theorem for the lattice TDDFT as well as a recent proof for the lattice TDCDFT \cite{TokatlyTDCDFT2011} essentially rely on the concept of the $v$-representability subset $\Omega$ in the Hilbert space. It would be very interesting and useful for the future to carefully study and to characterize the structure/geometry of $\Omega$. Formally the $v$-representability subset is defined as a part of $\mathcal{H}$ where ${\rm det}[\hat{K}]\ne 0$. Similarly to the explicit example of Sec.~IV, in general $\Omega$ should consist of two regions with different signs of ${\rm det}[\hat{K}]$, which are separated by a surface  ${\rm det}[\hat{K}]=0$. What is the geometry of each part for a general quantum system? Are they simply- or multiply-connected? Is it possible for system driven a physical potential to cross the surface ${\rm det}[\hat{K}]=0$ or it is forever confined to one initially fixed subregion of $\Omega$. Answering these questions will definitely deepen our understanding of TDDFT-related theories as well as quantum dynamics in general.

\begin{acknowledgments}
We are grateful to Angel Rubio, Stefan Kurth and David Cardamone for stimulating discussions and to Peter Schmitteckert for bringing the paper of Ref.~\onlinecite{SchDziSch2012} to our attention.

This work was supported by the Spanish MICINN (FIS2010-21282-C02-01) and ``Grupos Consolidados UPV/EHU del Gobierno
Vasco'' (IT-319-07).
\end{acknowledgments}

\appendix*
\section{Universal NLSE for two interacting particles on a two-site lattice (Hubbard dimer)}

Below we present the derivation of the universal NLSE for a system of two interacting spin-1/2 fermions on a two-site lattice. This gives another explicit example illustrating a general structure of nonlinear equations appearing in Sec.~III.

Since the dynamics of a triplet state on two-sites is trivial we concentrate on a singlet state. The spatial part of the singlet 2-particle wave function is symmetric with respect to the permutation of coordinates. It has three components, $\psi_{1 1}$, $\psi_{2 2}$, and $\psi_{12}=\psi_{21}$, describing different distribution of the particles over the sites. Because only the potentials orthogonal to a constant are relevant (see Sec.~IV) we assume from the very beginning that $v_1 = -v_2 =v$

The Schr\"odinger equation Eq.~\eqref{SE1} for two particles in a singlet state on a two-site lattice takes the form 
\begin{subequations}
 \label{2SE}
\begin{eqnarray}
i \partial_t \psi_{11}&=&-T \psi_{12} + (2v+w_{11})\psi_{11}\\
i \partial_t \psi_{12}&=& -T \psi_{11}+w_{12}\psi_{12} -T \psi_{22} , \label{2SEb}\\
i \partial_t \psi_{22}&=& -T \psi_{12}+(-2v+w_{22})\psi_{22}. \label{2SEc}
\end{eqnarray}
\end{subequations}

To construct the NLSE for this system we need to substitute the potential from the force balance equation which relates on-site potential to the density and the wave function. As we have a two-site lattice the structure of the force balance equation coincides with that of Eq.~\eqref{1LF} derived in Sec.~IV. We only have to substitute into Eq.~\eqref{1LF} the values of $k_{12}=2T\Re \rho_{12}$ and $q_{\bf r}$ calculated for the two particle system [by means of Eqs.~(\ref{rho1}) and (\ref{ST1})]  
\begin{equation}
k_{12}=2T\Re \rho_{12}=4 T \Re[\psi_{11}^*\psi_{12}+\psi_{12}^*\psi_{22}] ,
\end{equation}
\begin{eqnarray}
q_1=&-&4 T ^2 (|\psi_{22}|^2-|\psi_{11}|^2)+4T \Big( \Re[\psi_{11}^* \psi_{12}](w_{11}-w_{12})\nonumber  \\
&-&\Re [\psi_{22}^*\psi_{12}] (w_{22}-w_{12}) \Big) = -q_2 .
\end{eqnarray} 

As we know $\hat K$ matrix has a trivial zero eigenvalue corresponding to the constant potential which is already projected out by setting $v=v_1=-v_2$ . If $2T \Re \rho_{12} \neq 0$ we can invert $\hat K$ matrix in Eq.~\eqref{1LF}, which determines the potential $v$ in terms of density and the wave function 
\begin{equation} \label{deltav}
 v(n,\psi) = -\dfrac{\ddot n_1 + q_1}{4T \Re \rho_{12}}.
\end{equation}

Finally by substituting $v$ of Eq.~(\ref{deltav}) into Eq.~\eqref{2SE} we arrive at the universal NLSE for this system
\begin{subequations}
 \label{2NLSE}
\begin{eqnarray}
i \partial_t \psi_{11}&=&\Big(- \dfrac{\ddot n_1 + q_1}{2T \Re[\psi_{11}^*\psi_{12}+\psi_{12}^*\psi_{22}]}+w_{11}\Big)\psi_{11}
\nonumber \\
 &-&T \psi_{12}  ,  \label{2NLSEa}\\
i \partial_t \psi_{12}&=& w_{12}\psi_{12} -T \psi_{11} -T \psi_{22} ,  
\label{2NLSEb} \\
i \partial_t \psi_{22}&=& \Big( \dfrac{\ddot n_1 + q_1}{2T \Re[\psi_{11}^*\psi_{12}+\psi_{12}^*\psi_{22}]}+w_{22} \Big)\psi_{22} \nonumber \\
&-&T \psi_{12}. 
\label{2NLSEc}
\end{eqnarray}
\end{subequations}
We again explicitly see a system of nonlinear ODEs with the nonlinearity of the rational form. As one expects on the general grounds (see Sec~III) inclusion of interactions does not introduce any conceptual modification in comparison with the simplest one-particle case considered in Sec.~IV. In the subset $\Omega$ of the Hilbert space where $\Re \rho_{12}  \neq 0 $ all the terms in the right hand side of Eqs.~\eqref{2NLSEa} and \eqref{2NLSEc} are infinitely differentiable with respect to components $\psi_{\rv\rv'}$ of the wave function. Hence the whole right hand side stays Lipshitz continuous, provided $\psi\in \Omega$. If the density has a continuous second time derivative $\ddot n$, and the initial state $\psi(0)\in \Omega$ is density-consistent in a sense of Eqs.~\eqref{n_0-lat}--\eqref{dotn_0-lat}, then there is a unique solution to NLSE of Eqs.~\eqref{2NLSE}. Therefore there exists a one-to-one mapping between the density and the potential in the original two-body problem.

%
\end{document}